\documentclass[acmsmall,screen]{acmart}

\AtBeginDocument{%
  }

\setcopyright{acmlicensed}
\copyrightyear{2025}
\acmYear{2025}
\acmDOI{XXXXXXX.XXXXXXX}

\acmJournal{TIST}
\acmVolume{37}
\acmNumber{4}
\acmArticle{111}
\acmMonth{8}

\usepackage{listings}
\usepackage{xcolor}
\usepackage{longtable}
\usepackage{algorithm}
\usepackage{algpseudocode}
\usepackage{caption}
\usepackage{subcaption}
\usepackage{graphicx}
\usepackage{ulem}
\usepackage[flushleft]{threeparttable}

\lstdefinelanguage{json}{
    backgroundcolor=\color{gray!15},            
    basicstyle=\ttfamily\scriptsize,
    numbers=left,
    numberstyle=\tiny\color{gray},
    stepnumber=1,
    numbersep=8pt,
    numbers=none,
    showstringspaces=false,
    breaklines=true,
    frame=single,
    literate=
     *{0}{{{\color{blue}0}}}{1}
      {1}{{{\color{blue}1}}}{1}
      {2}{{{\color{blue}2}}}{1}
      {3}{{{\color{blue}3}}}{1}
      {4}{{{\color{blue}4}}}{1}
      {5}{{{\color{blue}5}}}{1}
      {6}{{{\color{blue}6}}}{1}
      {7}{{{\color{blue}7}}}{1}
      {8}{{{\color{blue}8}}}{1}
      {9}{{{\color{blue}9}}}{1}
      {:}{{{\color{red}{:}}}}{1}
      {,}{{{\color{red}{,}}}}{1}
      {"}{{{\color{black}{"}}}}{1}
}

\usepackage{array} 

\begin{document}

\title{Synthetic CVs To Build and Test Fairness-Aware Hiring Tools}

\author{Jorge Saldivar}
\email{jorge.saldivar@upf.edu}
\affiliation{%
  \institution{Universitat Pompeu Fabra}
  \city{Barcelona}
  \country{Spain}
}
\author{Anna Gatzioura}
\email{anna.gkatzioura@upf.edu}
\affiliation{%
  \institution{Universitat Pompeu Fabra}
  \city{Barcelona}
  \country{Spain}
}
\author{Carlos Castillo}
\email{carlos.castillo@upf.edu}
\affiliation{%
  \institution{Universitat Pompeu Fabra}
  \city{Barcelona}
  \country{Spain}
}








\renewcommand{\shortauthors}{Saldivar, Gatzioura and Castillo}

\begin{abstract}
Algorithmic hiring has become increasingly necessary in some sectors as it promises to deal with hundreds or even thousands of applicants. At the heart of these systems are algorithms designed to retrieve and rank candidate profiles, which are usually represented by Curricula Vitae (CVs). Research has shown, however, that such technologies can inadvertently introduce bias, leading to discrimination based on factors such as candidates' age, gender, or national origin. Developing methods to measure, mitigate, and explain bias in algorithmic hiring, as well as to evaluate and compare fairness techniques before deployment, requires sets of CVs that reflect the characteristics of people from diverse backgrounds. 

However, datasets of these characteristics that can be used to conduct this research do not exist. To address this limitation, this paper introduces an approach for building a synthetic dataset of CVs with features modeled on real materials collected through a data donation campaign. Additionally, the resulting dataset of 1,730 CVs is presented, which we envision as a potential benchmarking standard for research on algorithmic hiring discrimination.
\end{abstract}




\maketitle

\section{Introduction}\label{sec:introduction}

Although the European Union (EU) has long recognized equal employment as a force of social cohesion, dignity, and equality, its ambition remains unfulfilled as structural, institutional, and individual forms of discrimination continue to persist in the workplace \cite{chander2017}. The participation of women in the workforce continues to be unequal, and COVID-19 has derailed gender equality gains \cite{eige2021}. Similarly, ethnic minorities, people of African descent, women of color, and LGBTQ+ people have been shown to still be discriminated against.

In this context, algorithmic hiring is on the rise and rapidly becoming necessary in some sectors, as job postings that used to attract about 120 applicants in 2010 now attract over 250 \cite{fuller2021hidden}. AI technologies promise to deal with hundreds or thousands of applicants at high speeds. Moreover, their uptake in European HR teams and Public Employment Services is growing faster than the global average \cite{high2018high}. 

A key component in the realm of algorithmic hiring, i.e., the use of software technology to automate or assist employers through the different stages of recruitment (sourcing, screening, interviewing) \cite{li2021algorithmic}, is the algorithms to retrieve and rank candidate profiles, which are usually materialized in a curriculum vitae (CV). They are at the core of job search engines and applicant tracking systems (ATS), which are the tools that can lead to algorithmic discrimination. It is known that discrimination in hiring can be due to personal data such as age \cite{harris2023mitigating} and gender \cite{dastin2022amazon}, and due to sensitive attributes like national origin \cite{kovacheva2018growth}, or race \cite{chen2023ethics}.

Research on measuring, mitigating, and explaining bias in hiring processes assisted by algorithms, needs primarily real-world CVs that reflect the characteristics of people from diverse demographic backgrounds. However, the availability of sensitive information (i.e., sexual orientation, religion/belief, or ethnicity) in the published datasets is scarce or simply nonexistent \cite{fabris2023fairness}. Also, the vast majority of datasets used in algorithmic hiring research are not publicly available due to the sensitive information typically contained in CVs, or they are composed of synthetic materials fabricated using aggregated, incomplete, or artificially annotated demographic data. 

A way to advance the field can come from curating and publishing resources comprising synthetic CVs with i) representative job experience data; ii) diverse sensitive and demographic attributes; and iii) privacy guarantees for the data subjects involved. Moreover, the EU Artificial Intelligence Act (AI Act), in its article 10 (5.a), explicitly recommends the use of synthetic data as a primary option to avoid processing personal data when aiming to detect and mitigate algorithmic bias \cite{council2024regulation}.

In this context, the contributions of this paper are twofold: i) an approach to build synthetic CVs with characteristics resembling actual CVs; and ii) a dataset of 1,730 CVs, created using this approach, from a set of materials collected through a data donation campaign \cite{saldivar2025data}. We propose that this dataset could be used for benchmarking algorithmic hiring applications to promote increased fairness in ranking methods, especially when it comes to the evaluation and comparison against the state of the art of systems developed with relevance as a goal. 
Furthermore, we expect this dataset to become a standard reference for detecting biased ranking functions, in both academia and industry, while supporting increased explainability and research reproducibility.

The rest of the paper is structured as follows: Section \ref{sec:related_works} presents the theoretical foundation of our work and a review of the literature, while Section \ref{sec:method} introduces the proposed method, from the data collection to the generation of the dataset. In Section \ref{sec:results}, we present the results, starting with the characteristics of the synthetic dataset, followed by its validation. In Section \ref{sec:discussion}, we discuss our results, the main challenges faced and the elements we plan to improve in the future, to conclude our work in Section \ref{sec:conclusion}.
\section{Related Works}\label{sec:related_works}


In continuation, we present the theoretical background related to the main pillars of our work, namely: the most common application domain of synthetic data, their generation techniques, as well as some of the datasets that have been previously used in algorithmic hiring research.

\subsection{Synthetic data application domains}\label{sec:domains}

Computer vision \cite{paulin2023review}, and especially face recognition \cite{boutros2023face}, audio, natural language processing and health, have been presented as some of the main domains where synthetic data has had a significant impact \cite{hao2024synthetic}. Furthermore, various possible uses of synthetic data have been identified in human analysis \cite{joshi2024synthetic}, financial applications \cite{potluru2023synthetic}, as well as in human behavior and activity applications \cite{dahmen2019synsys}. 

Synthetic data has become a common approach for augmenting existing datasets or creating new ones, when real records are impossible to use due to logistic or ethical constraints \cite{whitney2024real}. In healthcare, for example, access to high quality data is particularly important, but also challenging due to privacy restrictions related to sensitive personal information and data sparsity, especially in cases of less common conditions \cite{murtaza2023health}. To this direction, synthetic data has been claimed to contribute to open research and innovation while enabling an improved diagnosis, treatment, and monitoring of diseases, being especially important for rare diseases or highly aggressive and complex conditions, to support more accurate, timely, and personalized healthcare solutions. The authors in \cite{rujas2025healthcare} present a comprehensive literature review and the uses of synthetic data in multiples healthcare domains (oncology, neurology, cardiology, etc.) and their subdomains. 

Synthetic data has been also widely recognized for its potential as a method to reduce bias in AI-based algorithms by removing imbalances and suppressing disparate impact, while maintaining data privacy, especially in high-risk applications, as in algorithmic hiring, referring to the use of systems that perform inferences for categorizing, classifying, ranking or recommending people \cite{kortylewski2019analyzing}.  
More details on the usage of synthetic data in algorithmic hiring are presented next in \ref{sec:algo_hiring}.

\subsection{Synthetic data generation techniques}\label{sec:techniques}

Synthetic data refers to artificially generated data that mimics the characteristics and patterns of real-world data, but is created through algorithms or generative models rather than being directly collected or annotated by humans \cite{liu2024best}. It relates to, but differs from, synthetic content generation (text, images, video, media, etc.), since rather than being created to be consumed by end-users, synthetic data are generated with the aim to be used as inputs to other data processing systems,  where the original data cannot be used due to size or privacy concerns \cite{critical2024}.
 
The first approaches to synthetic data generation and data imputation were based on non-parametric machine learning models, such as classification and regression trees, support vector machines and random forests, aiming to generate data based on the statistical distribution of attributes in real data \cite{nowok2016synthpop}. However, the recent advances in generative AI have caused a shift in the computational paradigms used in many applications, including data generation techniques. Generative AI models refer to AI methods that learn the underlying distributions in existing data and generate novel data according to them. Among those, variational autoencoders (VAEs) and generative adversarial networks (GANs) have been widely used to generate synthetic data \cite{jordon2022what}. 
As the generation process depends on the type of data to be generated, for text data also NLP, BERT and transformer models, like GPT, have been applied, while convolutional neural networks (CNNs) have been found to better learn image representations. Finally, in domains where the data sequence is important, recurrent neural networks (RNNs) are used to generate meaningful synthetic sequences \citep{lu2023review,eigenschink2023deep}.


To be considered as ``good'', synthetic data must be representative of the original data, while also providing guarantees about privacy \cite{howe2017}. Furthermore, the quality of synthetic data is highly dependent on both the quality of the original data and the generative model used: if the original data contains errors, biases, or under-represents certain features, the synthetic dataset will also reflect those flaws. Similarly, if the model is not able to properly capture the key characteristics of the original data, the generated data may not provide accurate insights. To this direction, Endres et al. highlight the importance of proper data collection and cleaning, as prior steps to training models to generate synthetic data \cite{endres2022}.

For a synthetic dataset to be useful in a given application domain, it should have multivariate distributional properties similar to the dataset based on which it was made, ensuring that it could substitute the original dataset with very similar analytic results. Furthermore, the intended use has to be defined in advance, to properly select the most adequate generation technique and take into account domain specific characteristics and limitations \cite{leveraging}.

An important aspect when generating synthetic data, and one of the main decisions to be made when building a model, is the desired trade-off between control and flexibility. Although some black-box approaches manage to produce high-dimensionality data and at a large scale, they are particularly opaque, making it hard to have control over the process and estimate their privacy. Therefore, their use is not adequate for applications domains like algorithmic hiring \cite{kapania2025examining}. As our research scope has been to generate a data set of synthetic CVs that could be used as a benchmarking dataset, under the premise that it would reflect the distribution of the original characteristics, while preserving anonymity, we have opted to develop a hybrid technique that allows us to have more control over the process and, as a consequence, a better understanding of the results.

\subsection{Datasets and synthetic data in algorithmic hiring research}\label{sec:algo_hiring}

The study of fairness in algorithmic hiring requires access to personal data about job applicants, including their work experience and sensitive attributes. Unfortunately, there is a lack of publicly available resources comprising the CVs of candidates and their protected attributes \citep{fabris2023fairness}. Previous work on fairness in CV-based hiring has used private databases \citep{cowgill2018bias,parasurama2022gendered} or short text snippets with limited sensitive attribute information \citep{de2019bias,zhang2022male}. 

To highlight the importance and challenging aspects of algorithmic tools for job discovery and hiring, the annual RecSys challenge\footnote{Please refer to RecSys 2016: \url{http://2016.recsyschallenge.com}; RecSys 2017: \url{https://www.recsyschallenge.com/2017}} focused on job recommendations both in 2016 \cite{RecSysChallenge2016_TrainingDataset} and 2017 \cite{RecSysChallenge2017_Dataset}. A data set from the XING platform\footnote{Please refer to \url{https://www.xing.com}} was used. This dataset was formed on a semi-synthetic resumes, 
enriched with noise to anonymize and abstract from real user profiles. However, the aim of both years' tasks has been different from our purpose, as this 
dataset served for building models to determine the relevance and potential interest of users to job postings under different use cases.

Our research scope is quite different from the majority of works with synthetic data in the recruitment domain, as rather than exploring the linguistic characteristics appearing in online CVs \cite{decorte2021jobbert}, our goal is the creation of a dataset of synthetic CVs, following the distributions of the characteristics in donated data, to enable algorithmic fairness benchmarking.
In contrast, the majority of CV or biography datasets are compiled as supportive elements for classification or linguistic exploration tasks rather than to improve research related to fairness in algorithmic hiring. 

For example, Skondras et al. explore the potential of large language models to augment a dataset of real resumes crawled from Indeed\footnote{Please refer to \url{https://www.indeed.com}} with artificially generated resumes created with chatGPT. Resumes in dataset belong to 15 different job categories, ranging from technical roles, like software developers, to professional roles, such as lawyers. The authors' main goal was to test whether the augmented dataset helps to improve resume classification tasks \cite{skondras2023generating}. 

In \cite{jiechieu2021skills} the authors also used Indeed to construct a corpus based on approximately 28k anonymous IT resumes being available on this platform. Their objective was the identification of skills relevant to job roles based on the unstructured textual information found in resumes and the classification and relevance calculation of those, rather than the use of this corpus to support the generation of synthetic curricula vitae.
One of the few open datasets found is the Kaggle ``Resume dataset'' \cite{kaggle_dataset} which includes around 24k resumes taken from LiveCareer\footnote{Please refer to \url{https://www.livecareer.com}}, in text format, and the most suitable profession for each, from 25 distinct profession sectors (e.g., Designer, IT, Teacher, Business, Healthcare, Agriculture), aiming to support an improved categorization of resumes into defined labels. \textcolor{black}{Another open dataset is presented in \cite{drushchak-romanyshyn-2024-introducing}, where the authors introduce a collection of 230k real CVs extracted from the Ukranian platform Djinni\footnote{Please refer to \url{https://djinni.co}}, primarily in English and from the IT sector, which also includes within the content of the documents explicit mentions to some demographic attributes such as age, gender, marital and military status, and religion}.

Peña et al. \cite{pena2020bias} study how current multi-modal algorithms for algorithmic hiring are affected by sensitive attributes and preexisting biases in the data. They use a dataset of 24k synthetic resumes covering 4 job sectors, with 12 features including education, availability, previous experience, occupation, name, and language, 2 demographic attributes (gender and ethnicity) and a face photograph from the DiveFace database \cite{morales2020sensitivenets} correlated with the demographic attributes. They further extended this dataset \cite{pena2023human} by including short biographies to the CVs using the Common Crawl Bios dataset \cite{de2019bias} which contains online biographies related to 28 different occupations. The Common Crawl Bios dataset \cite{de2019bias} was created with the aim to study gender bias in occupations' classification and consists of almost 400k biographies. Although in both works bias in algorithmic hiring is considered an important aspect, their focus is on its detection, rather than on the generated datasets. In addition, in the synthetic CV dataset that we propose, only structured information related to education, work experience and obtained skills, is included. We do not assign images or names to the generated CVs.

Bruera et al. \cite{bruera2022generating}, starting from the Indeed dataset used by \cite{jiechieu2021skills}, propose a method that combines Bayesian networks and natural language processing techniques to resemble candidate attributes as found in a dataset of real CVs. They first get the initial structure of a CV and then complete it using prompts to query a generative model. In contrast to our work, they include artificial personal information in their synthetic CVs. This work is probably the closest to ours; however, the authors aim to generate a synthetic dataset permitting to train machine learning models and do not focus on fairness issues or the benchmarking of ranking algorithmic in terms of fairness. Table \ref{tab_previous-datasets} summarizes the information from the CV datasets presented in this section. 

\begin{table}[h]
    \begin{threeparttable}
        \caption{Datasets of CVs published in the literature}
        \label{tab_previous-datasets}
        \scriptsize
        \begin{tabular}{p{0.3cm}p{1cm}>{\raggedright}p{1.3cm}p{0.8cm}p{1.1cm}>{\raggedright}p{4cm}p{1.5cm}p{1cm}}\toprule
        Ref. & Type & Source(s) & N & Job \mbox{sectors} & Main contents & Demographics & Availability \\\midrule
        \cite{RecSysChallenge2016_TrainingDataset} & Semi-synthetic & Xing & Unknown & Unknown & Education, experience, career level, \mbox{discipline,} industry, country & No & Unavailable \\
        \cite{RecSysChallenge2017_Dataset} & Semi-synthetic & Xing & Unknown & Unknown & Education, experience, career level, \mbox{discipline,} industry, country & No & Unavailable \\
        \cite{skondras2023generating} & Hybrid (Real, synthetic) & Indeed, \mbox{ChatGPT} & 2K & 15 & Unknown & No & Upon \mbox{request} \\
        \cite{jiechieu2021skills} & Real & Indeed & 28K & 1 & Unknown & No & Unavailable \\
        \cite{kaggle_dataset} & Real & Livecareer & 24K & 25 & Education, experience, skills & No & Public \\
        \cite{drushchak-romanyshyn-2024-introducing} & Real & Djinni & 230K & 1 & Education, experience, skills, english level, years of experience, name & Age, gender, religion, marital and military status & Public \\
        \cite{pena2020bias} & Synthetic & DiveFace, US~Census Bureau 2018 & 24K & 4 & Education, occupation, availability, \mbox{experience,} languages, name, photo & Gender, \mbox{ethnicity} & Unavailable \\
        \cite{de2019bias} & Real & Common Crawl & 400K & 28 & Short biography, name & Gender & Unavailable \\
        \cite{bruera2022generating} & Synthetic & \cite{jiechieu2021skills} & 4K & 1 & Skills, education, experience, hobbies & No & Unavailable \\
        \bottomrule
        \end{tabular}
    \begin{tablenotes}
      \footnotesize
      \item \textbf{Note.} \textit{Ref.}: reference; \textit{N}: number of CVs in the dataset; \textit{Jobs sectors}: number of job sectors represented in the dataset. The dataset \cite{pena2023human} reviewed above was not considered in the table, as it is an extension of \cite{pena2020bias}, which is already included.
    \end{tablenotes}
    \end{threeparttable}
\end{table}

\section{Method}\label{sec:method}

Our proposed approach is to generate a dataset of synthetic CVs based on a reference dataset built from data donated by consenting individuals whose identities were safeguarded by anonymizing the donated data.

\subsection{Data Donation Campaign}\label{sec_data_donation}

We proposed collecting real CVs and attributes potentially leading to discrimination (i.e., age, gender, religion, origin, and disability condition) through a data collection campaign. Residents of the European Economic Area and Switzerland who are part of the labor force (i.e., employed or seeking employment) were invited to voluntarily donate up to two anonymous CVs by completing an online survey\footnote{Please refer to \url{https://findhr.eu/datadonation} 
to access the website used to run the campaign.}. Donors were recruited online and provided with an information sheet and a consent form \cite{saldivar2025data}.

The campaign started in June 2023 and remained open until the end of May 2024. In total 1,143 donations were received, four of which were discarded because of incomplete fields. The remaining 1,139 completed submissions included 1,211 CVs, considering that about 15\% of them contained two CVs (up to two CVs could be attached to the submission). Most donations were in Spanish (78\%, 895 out of 1,139), impacting the language of CVs, which are primarily written in Spanish (69\%, 836 out of 1,211).

Half of the donors declared themselves as professionals (567 out 1,139) from various sectors, including science and engineering, business and administration, ICT, legal, social, cultural, and health. An interesting balance between junior and senior workers is available in the data. 
Fifty percent of the donors are between 26 and 45 years of age (581 out of 1,139), and half of donations were submitted by women (576 out of 1,139). Almost 20\% of donors reported belonging to the LGBTQ+ community (192 out of 1,139), a similar proportion declared being part of a minority group (229 out of 1,139), and 15\% perceived themselves as foreign in the country where they live (179 out of 1,139). About 45\% of the campaign participants (491 out of 1,139) reported being either secular or not religious; the rest are mostly Christians, while Muslims, Buddhists, Hinduists, and Jews are marginally represented in the data. Less than 10\% of donors (84 out of 1,139) declared having a disability condition.

Although most donors are Spanish nationals, the donation dataset is representative of the European workforce in terms of gender, age, and foreign condition (i.e., whether donors feel foreign in the country where they live) \cite{saldivar2025data}. On the other hand, it overrepresents people from the LGBTQ+ collective, ethnic minorities, and disabled communities, which is somewhat expected given the evidence of employment discrimination suffered by these groups \cite{yam2021human,bertrand2004emily,tilmes2022disability}. Regarding religion, Europe is majority Christians \cite{EU2019_HLG_DigitalTransformation}, while donations were primarily submitted by agnostic/secular people. The group of Muslims is, however, fairly well represented, whereas other religions, like Buddhists, Jews, and Hinduists, are represented in both the general population and donations with less than 1\%.

Some professional sectors, such as ICT, engineering, clerical support, and business administration, are overrepresented in the donations. We understand that this is related to the digital means used to run and advertise the campaign, which might facilitate donations from these sectors, i.e., those using computers to support their daily tasks, but complicates participation from sectors that typically involve manual labor. Another explanation might be that overrepresented sectors align with professionals familiar with using CVs and job search platforms to access the labor market.

The collected sample with the above characteristics is used as a reference for constructing the synthetic data. We aim to generate synthetic CVs with statistical properties resembling those of people who participated in the donation campaign.

\subsection{Synthetic CV Generation Approach}

\begin{figure}[tb]
  \centering
  \includegraphics[width=1\textwidth]{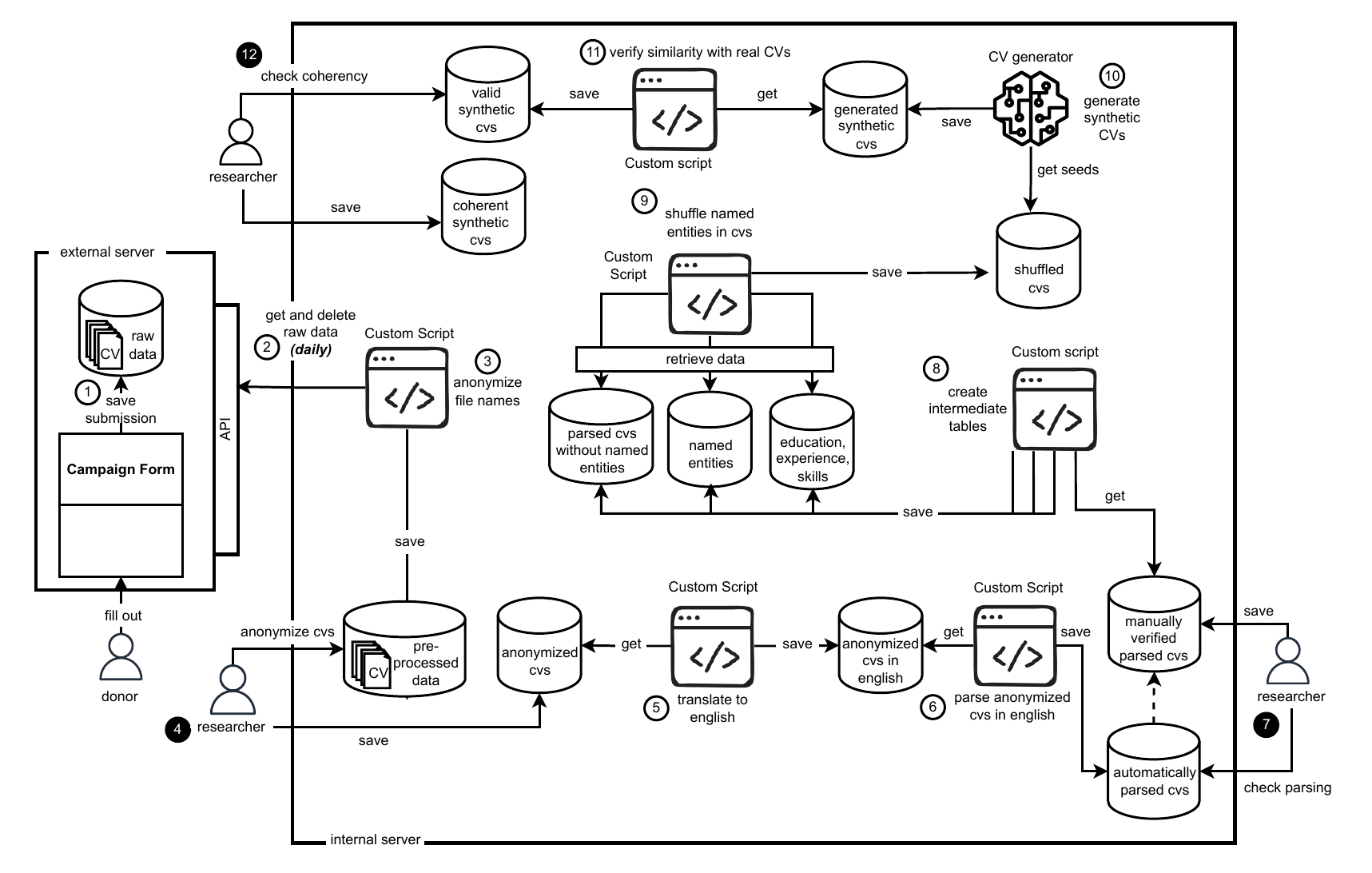}
  \caption{Pipeline proposed to generate the synthetic dataset. Black circles indicate manual tasks.}
  \label{fig_pipeline}
\end{figure}

We propose an approach that combines automatic and manual tasks to generate the synthetic data. Figure \ref{fig_pipeline} shows the pipeline to collect, process, store, and produce the synthetic CVs. For our purpose, a synthetic CV is defined as a structured text document organized in three sections: \textit{educational background}, \textit{professional experience}, and \textit{skills}, each containing a number of items representing information related to the section.

The pipeline starts by processing the donations as they arrive (step 1 in Figure \ref{fig_pipeline}). This involves anonymizing the CV file name(s), creating a directory to allocate the CVs, and generating a JSON file containing all submission data as is, i.e., without any further processing (steps 2 and 3). In step 4, donated CVs are manually anonymized, removing all personally identifiable information (names, addresses, pictures, emails, phone numbers). Later, CVs written in languages different from English are automatically translated into English (step 5).

Anonymized CVs in English and formatted as PDFs are parsed by hitting the RESTful API of the GDPR-compliant resume parsing service EdenAI (step 6). EdenAI\footnote{Please refer to \url{https://www.edenai.co}} was chosen as the parser due to the time, cost, and human resources limitations of building an in-house parsing tool. The parsing is automatically processed, generating the JSON structure shown in Listing \ref{example-json-structure}. Later, the created JSON is manually verified (step 7), filling out missing information or fixing incorrect data. At this step, the processing of CVs finishes, and what remains is related to the generation of synthetic CVs.

\begin{lstlisting}[language=json, label=example-json-structure, caption={Illustrative example of a JSON that resulted from processing the parsing output}, float]
{
    "education_background": [
        {
            "degree": "BSc. Computer Science",
            "start_date": "2019-09-15",
            "end_date": "2023-06-30",
            "institution": "Pompeu Fabra University"
        },
        { ... }
    ],
    "professional_experience": [
        {
            "role": "Junior Software Developer",
            "start_date": "2023-01-01",
            "end_date": "Present",
            "institution": "AI company",
            "description": "..."
        }
    ],
    "skills": [
        "hard": ["Python", "MySQL", ...],
        "soft": ["Leadership", "Teamwork", ...],
        "languages": ["English", "Spanish", ...],
        "others": ["Photography", "Public speaking", ....]
    ]
}
\end{lstlisting}

\subsubsection{Intermediate dataset} \label{intermediate_dataset}

In preparing the generation and as a way to reduce potential re-identification risks, information in the parsed CVs is extracted from JSON files and split into three unrelated tables (step 8). One of the tables, \textit{anonymized-cvs}, contains, in random order, one record per anonymous, parsed, and processed CV in JSON format, as well as the job sector, years of professional experience, and demographic data (age, gender, origin, etc.) associated with the CV (see Table \ref{tab_anonymized-cvs}). CVs included in this table not only miss personally identifiable information, but also names of institutions (companies, universities, NGOs) are removed from the education and professional items. Due to parsing inconsistencies, education dates may not reflect typical study durations. To address this, rules--such as a minimum of three years for Ph.D. programs or nine months for master's--are applied before adding entries to the anonymized table. The other table, \textit{education-experience-skills-combinations}, involves, in random order, one record per CV, including the job sector and experience, together with the list of educational institutions, workplaces, and skills contained in the donated CV (see Table \ref{tab_edu_exp_skills}). The last table, \textit{named-entities}, contains a job sector, a value of a sensitive variable (e.g., ``Woman'' or ``Muslim''), the number of CVs in the job sector, including that value, which must be larger than or equal to 5 to be included, and an alphabetic list of entities found in CVs with that value, i.e., educational institutions and workplaces (see Table \ref{tab_named_entities}). The latter rule---included in the approved protocol for data generation (see section \ref{sec_ethical_consideration})---was arbitrarily decided to protect less-represented groups of donors.

\begin{table}[]
\begin{threeparttable}
\caption{Table anonymized-cvs 
}
\footnotesize
\begin{tabular}{p{3cm}p{1.5cm}p{3.5cm}p{1cm}p{1cm}p{1cm}p{0.2cm}}
\toprule
Anonymized and parsed CV & Job Sector & Year of professional experience & Age & Gender & Ethnicity & … \\ \midrule
\{...\} & IT & 12 & 38 & Woman & … & … \\
\{...\} & Engineering & 1 & 25 & Woman & … & … \\
\{...\} & Business & 7 & 31 & Man & … & … \\
... & ... & ... & ... & ... & ... & ... \\ 
\bottomrule
\end{tabular}
    \begin{tablenotes}
      \footnotesize
      \item \textbf{Note.} Illustrative example of the table containing, in random order, one record per anonymized and parsed CV plus the job sector, years of professional experience, and additional data (regular, sensitive). 
    \end{tablenotes}
\label{tab_anonymized-cvs}
\end{threeparttable}
\end{table}

These tables are created to protect donors by separating sensitive information from their CVs, feeding the generator with this intermediate dataset. Donated materials are never directly used to generate synthetic CVs. Also, the content of these tables is employed to add variability to the seed data by shuffling the education institutions and workplaces of donors with similar demographic characteristics (step 9). As part of this step, mappings between companies and professional roles and between education degrees and academic institutions are generated using Sentence-BERT \cite{reimers2019sentence}, a variant of BERT \cite{kenton2019bert} that uses siamese neural networks for large-scale text comparison. A manual inspection of the generated mappings demonstrates useful and reliable results.

Similarly, the probability distribution of skills associated with the degrees and roles is calculated.
In the process, a higher abstraction level is introduced to group together degrees, or roles, that refer to the same concept while different names have been used in the donated CVs (e.g., a bachelor's degree that has been introduced as ``bachelor in science'' and ``BSc''). This classification permits the calculation of probability distributions in relation to more general, educational, and working experience categories, leading to reduced data sparsity. The relevance of degrees and roles to skills is then stored into separate tables that contain, respectively, the educational degrees and their relevance to each of the observed skills and the job roles and their relevance to each of the skills.

\begin{table}[]
\begin{threeparttable}
\caption{Table education-experience-skills-combinations 
}
\scriptsize
\begin{tabular}{p{1.5cm}p{2cm}p{3cm}p{2cm}>{\raggedright\arraybackslash}p{3.5cm}}
\toprule
Job Sector & Years of professional experience & Educational institutions & Workplaces & Skills \\ \midrule
IT & 12 & Pompeu Fabra University, ... & AI Company, ... & Python, MySQL, Leadership, Teamwork, English, Spanish, Photography, Public speaking \\
Engineering & 1 & Complutense University of Madrid, ... & Sakyr SA, ... & Calculus, Excel, AutoCAD, Teamwork, Portuguese, Italian, Running, Communication \\
Business & 7 & INCAE Costa Rica, ... & Inditex, ... & Excel, PowerBI, Finance, Leadership, Spanish, Italian, Guitar \\
... & ... & ... & ... & ... \\ 
\bottomrule
\end{tabular}
\label{tab_edu_exp_skills}
    \begin{tablenotes}
      \footnotesize
      \item \textbf{Note.} Illustrative example of the table containing, in random order, one record per donated CV with job sectors and years of professional experience, list of educational institutions of an individual, list of workplaces of the individual, and words/phrases describing the individual’s skills. 
    \end{tablenotes}
\end{threeparttable}
\end{table}

\begin{table}[]
\begin{threeparttable}
\caption{Table named-entities 
}
\scriptsize
\begin{tabular}{p{1.5cm}p{2cm}p{1cm}p{1cm}p{1cm}>{\raggedright\arraybackslash}p{3cm}p{2cm}}
\toprule
Job Sector &  Year of professional experience & Variable &  Variable Value & Num. CVs &  Education Institutions & Workplaces \\ \midrule
IT & 12 & Age & 22 & 37 &  Pompeu Fabra University, ... & AI Company, ... \\
IT & 12 & Gender & Woman & 18 &  Pompeu Fabra University, ... & AI Company, ... \\
Engineering & 1 & Age & 25 & 31 &  Complutense University of Madrid, ... & Sakyr SA, ... \\
Engineering & 1 & Gender & Woman & 12 &  Complutense University of Madrid, ... & Sakyr SA, ... \\
Business & 7 & Age & 31 & 43 &  INCAE Costa Rica, ... & Inditex, ... \\
Business & 7 & Gender & Man & 23 &  INCAE Costa Rica & Inditex, ... \\
... & ... & ... & ... & ... & ... & ... \\ 
\bottomrule
\end{tabular}
\label{tab_named_entities}
    \begin{tablenotes}
      \footnotesize
      \item \textbf{Note.} Illustrative example of the table containing a job sector, a value in a regular or sensitive category data, the number of CVs in the job sector including that value, which must be larger than or equal to 5 to be included, and an alphabetic list of entities found in CVs with that value, i.e., locations, educational institutions, and workplaces. 
    \end{tablenotes}
\end{threeparttable}
\end{table}

\subsubsection{Final steps of the process}\label{final_steps_generation}

After the generation (step 10), synthetic CVs are automatically verified (step 11) to ensure they do not closely resemble the donated materials of any individual donor. The final step (step 12) involved evaluating the generated CVs. A research assistant —trained by the authors— manually reviewed each synthetic CV, assessing its coherence with the target professional sector. The guiding question was whether the education, work experience, and skills described in the CV reflect a plausible career trajectory within that sector. For example, a full-stack developer with a bachelor degree in software engineering would not be coherent with the \textit{clerical support} sector, aligning instead with \textit{ICT}.

In addition, the coherence within and between sections (education, experience, skills) is evaluated, allowing flexibility for non-traditional career paths. This means that a political science graduate who worked in customer support at a telecommunications company while studying would be considered coherent, whereas a graduate in web design and marketing working as a motorcycle mechanic would not. Similarly, it is implausible for a seller with a high-school diploma to be presented as an expert in schizophrenia therapies. Moreover, highly divergent professional or academic trajectories, such as completing a master's degree in microprocessors followed by a PhD in law, or serving as a CEO before working as an accounting assistant, while possible, are not included in our resulting dataset. Each criterion was rated on a scale from 1 (poor) to 5 (excellent). After this process, only CVs with an overall rating of at least 4 (very good) out of 5 were retained; the rest were discarded. 


\subsection{Synthetic CV Generator}\label{sec_synthetic_cv_generator}

The primary goal of the synthetic CV generator (step 10 in Figure \ref{fig_pipeline}) is to resemble the characteristics of real donated CVs. Generated documents do not contain personal information (e.g., name, email address, phone number, or picture) 
and are structured into three sections, namely educational background, professional experience, and skills, each section containing a list of items.

The generation is based on the mandatory parameters: job sector and years of professional experience. It is required to specify the job sector of the CVs to be generated, as well as the years of professional experience that the synthetic CV should reflect. Also, at least one personal and/or sensitive attribute, like age, gender, disability condition, or ethnicity, should be specified. Next, the generator finds out what a ``typical'' CV of people with the specified characteristics looks like, i.e., it needs to recognize the typical characteristics of CVs belonging, for example, non-European female lawyers in their 30s who have 5 years of professional experience. In other words, it determines how to fill in the sections on educational background, professional experience, and skills with respect to the number of items in each section and their content.


\subsubsection{Structure Computation}

Following the example above, say we want to compute the number of items in the skills sections of a ``typical'' CV of non-European female lawyers in their 30s with five years of professional experience. We extract all anonymous CVs of non-European donors (from Table \ref{tab_anonymized-cvs}) and check the number of items in their skill sections. Then, a Weibull distribution \cite{rinne2008weibull} that satisfies these numbers is computed. A similar procedure is applied to the rest of the input parameters, i.e., in the example, gender (female), professional sector (law), age (30), and professional experience (five years).  Weibull distribution was chosen because sampling from this curve does not result in negative values, which is crucial in our case. Also, Weibull is typically used in survival analysis, and hiring can be seen as a process in which candidates kind of ``strive to survive'' until they reach the final stages. Moreover, Weibull is shown to be sufficiently flexible despite requiring only two parameters (shape and scale).

At the end of this process, we have as many distributions as the number of input parameters. Each distribution is sampled, and the results are combined using a randomly chosen strategy, including mean, median, min, and max. The number of items for the section educational background and professional experience is calculated using this approach. For the skills section, the number of items is capped at 12, as generation tests showed that the section coherence drops significantly beyond this limit. Algorithm \ref{alg_items} shows the pseudo-code proposed to compute the number of items to be generated for each section.

\begin{algorithm}
\caption{Compute number of items per section}\label{alg_items}
\scriptsize
\begin{algorithmic}[1]
\Require $input\_parameters$ \Comment{Array of input parameters}
\State $items \gets \Call{Init\_Dictionary}{ }$ \Comment{Initialize an empty dictionary}
\State $num\_params \gets \Call{Length}{input\_parameters}$ \Comment{Return the length of the array as integer}
\For{$section \in \{education, experience, skills\}$}
\State $dists \gets \Call{Init\_Array}{num\_params}$ \Comment{Initialize an empty array of n elements}
\For{$i \in \{1, \dots, \textit{num\_params}\}$} 
\State $cvs \gets \Call{Extract\_CVs}{table\_\ref{tab_anonymized-cvs}, input\_parameters_i}$ \Comment{Get items in Table \ref{tab_anonymized-cvs}}
\State $num\_cvs \gets \Call{Length}{cvs}$
\State $num\_items \gets \Call{Init\_Array}{num\_cvs}$
\For{$j \in \{1, \dots, \textit{num\_cvs}\}$}
\State $num\_items_j \gets \Call{Get\_Num\_Items}{cv_j, section}$ \Comment{Get number of items in section}
\EndFor
\State $dists_i \gets \Call{Weibull}{\textit{num\_items}}$ \Comment{Draw Weibull distribution}
\EndFor
\State $samples \gets \Call{Init\_Array}{num\_params}$
\For{$k \in \{1, \dots, \textit{num\_params}\}$}
\State $samples_k \gets \Call{Sampling}{\textit{dists}_k}$  \Comment{Sample Weilbull distribution}
\EndFor
\State $samples\_comb \gets \Call{Mean}{samples}$ \Comment{Combine samples}
\State $items[section] \gets \Call{Round}{samples\_comb, 0}$ \Comment{Round value to 0 decimals}
\EndFor
\State \Return $items$
\end{algorithmic}
\end{algorithm}

\subsubsection{Content Generation}\label{sec_content_generation}
Once the number of items in each section is known, the next step involves deciding how to fill them, and here is where the tables created during the processing tasks (step 8 in Figure \ref{fig_pipeline}) come in handy. The procedure proposed to create the content of synthetic CVs, 
is described next.

\begin{enumerate}
    \item Collect CVs that satisfy the input parameters by querying Table \ref{tab_anonymized-cvs}. Remember that CVs included in this table are in JSON format, do not contain personal information, and miss the names of all institutions reported in the document. Going back to the example, this means recovering all CVs of non-European female lawyers in their 30s with five years of professional experience. The number of CVs must be larger than 20 to continue; otherwise, the process is abandoned. This rule was decided based on a sufficiently large number of CVs (20) to avoid compromising donors' identity and is part of the safeguard protocol approved by the Institutional Committee for Ethical Review of Projects at Pompeu Fabra University, see section \ref{sec_ethical_consideration};
    \item Use Table \ref{tab_named_entities} to extract institutions (universities and workplaces) belonging to CVs that satisfy the input parameters. That is to say, create a list with the institutions contained in the CVs of non-European lawyers with five years of professional experience plus the institutions available in CVs of female lawyers with five years of professional experience plus institutions in CVs of lawyers in their 30s with five years of professional experience;
    \item Fill in the missing university and workplace names of CVs collected before with those extracted in the previous step. The mappings created in step 9 are employed here to ensure coherent replacements. The chosen institutions are checked against the Table \ref{tab_edu_exp_skills} to ensure they do not accidentally match or near-match (match, except for one parameter) a real combination for a donor;
    \item Separate CVs' items into three groups: education, experience, and skills. Compute embeddings for the items in each group and cluster the embeddings using Agglomerative Clustering \cite{taha2023semi};
    \item Generate sections education background, professional experience, and skills by selecting the required number of items according to the following generative procedures:
    \begin{itemize}
        \item For the \textbf{education background} section, it is arbitrarily defined that it can have, at maximum, five items, namely, one Bachelor, one abroad experience (e.g., Erasmus, internships, visiting scholar periods), two Masters, and one PhD. The procedure for the education background section starts by randomly selecting a bachelor item---if, for the given combination of parameters, reference CVs do not contain bachelor items that can be selected, a vocational experience (if any) is chosen, and the process stops. Next, at maximum, one abroad, two Masters, and one PhD experience that belong to the same cluster of the picked bachelor are randomly selected. Whether to include or not these items is guided by a stochastic mechanism that aims to increase diversity in the produced education sections. If an abroad experience is decided to be included, it should be within the period of the already included bachelor, Master, or PhD. If master experiences are picked, their start dates should be later than the end date of the included bachelor. A similar logic is applied to a PhD, meaning if a PhD experience is chosen, it cannot start before the Master's ends;
        \item For the \textbf{professional experience} section, items are added subsequently, ensuring that picked items follow a chronological order. In the worst case, if no professional experience comes later than the initially selected item, the section has only one item. But, on average, between one and the required number of professional experience items are included. Another restriction imposed on the procedure is that the total duration of the selected professional experiences cannot surpass the expected professional experience specified in the mandatory parameter \textit{years of professional experience}. Like the procedure to generate the education section, the process starts by randomly selecting a professional experience. If the duration of the chosen experience is longer than the expected \textit{years of professional experience}, the process stops. Otherwise, the process continues by randomly selecting subsequent items that belong to the cluster of the first chosen experience while the duration of the selected professional experiences altogether does not exceed the specified \textit{years of professional experience};
        \item For the \textbf{skills} section, we identify the skills being most relevant to the items that are included in the education background and professional experience sections. 
        The selection is based on probability distributions (calculated during the pre-processing phase, see \ref{intermediate_dataset}) that associate education backgrounds and working experiences with skills.
        As skills may be associated with 
        multiple education and professional entries, the retrieved distributions are aggregated and ranked. Finally, the top \textit{n} skills are selected, where \textit{n} represents the number of items that were previously calculated to be included in the skill section. In cases where a smaller number of relevant skills are identified, all of them are included.
    \end{itemize}
\end{enumerate}

\subsubsection{Generation Rules}
A set of rules has complemented the generation procedure to guarantee a certain quality in the resulting fabricated CVs and minimize clumsy inconsistencies and errors. Generally, there cannot be two similar synthetic CVs in a generation attempt. Second, items either in the skills section as well as in the educational background or professional experience cannot be duplicated. Synthetic CVs with empty sections are rejected. 

\subsubsection{Generation Parameters}

In generating the synthetic CVs, we aim to resemble the characteristics of the donated materials. To satisfy this requirement, we need to feed the generator with some parameters, including both professionally and personally related information. 

Mandatory parameters correspond to the job sector and years of professional experience that CVs should reflect. Values that the mandatory parameters can take are presented in Table \ref{tab_mandatory_params}. Names of all job sectors, except Public officials, correspond to categories of the European classification of Skills, Competences and Occupations (ESCO), which were used in the donation form (see Section \ref{sec_data_donation})\footnote{Interested readers can check ESCO website (\url{https://esco.ec.europa.eu/en/classification/occupation}) for information about the meaning of each job sector}. Public officials, on the other hand, is a renaming of the ESCO category \textit{Chief executives, senior officials and legislators} aiming to look for a more concise and representative term that covers all professional activities in this sector.

\begin{table}[]
\caption{Mandatory generation parameters and their corresponding values}
\label{tab_mandatory_params}
\scriptsize
\begin{tabular}{p{1.5cm}>{\raggedright\arraybackslash}p{7cm}>{\raggedright\arraybackslash}p{4cm}}
\toprule
Parameter & Job sector & Years of professional experience \\
\midrule
Values & Business and administration; Clerical support; ICT; Science and engineering; Sales; Legal, social and cultural; Construction, manufacturing and transport; Health; Public officials; Production and specialized services; Personal service; Teaching; Cleaning; Food preparation; Food Processing, Woodworking, Garment and Other Craft; Agricultural, forestry and fishery; Armed forces; Hospitality, retail and other services; Personal care; Handicraft and Printing; Protective & 4 years or less, 5-9 years, 10-14 years, 15+ years \\ 
\bottomrule
\end{tabular}
\end{table}

In addition, the generator requires one or more demographic attributes, such as age, gender, ethnicity, or religion. Unlike the others, the attribute ethnicity was not operationalized through direct options but implemented via the indirect parameters \textit{Perceived foreign in the country of residence} and \textit{Belong to an ethnic minority in the country of residence}. This corresponds to how information about donors' ethnicity was collected. Table \ref{tab_optional_params} lists the parameters related to the demographic attributes.

\begin{table}[]
\caption{Generation parameters related to demographic attributes and their values}
\label{tab_optional_params}
\scriptsize
\begin{tabular}{p{1.5cm}>{\raggedright\arraybackslash}p{1cm}p{2cm}p{1cm}p{1cm}p{1cm}>{\raggedright\arraybackslash}p{2.5cm}p{1cm}}
\toprule
Parameter & Age & Disability condition & Gender & Minority & Perceived foreign & Religion & LGBTQ+ \\ 
\midrule
Values & $<=30$, 31-40, 41-50, $>50$ & Yes, No & Woman, Man, Non-binary & Yes, No & Yes, No & Buddhism, Christianity, Hinduism, Muslim, Judaism, Other, Secular & Yes, No \\ 
\bottomrule
\end{tabular}
\end{table}


\subsection{Ethical Considerations}\label{sec_ethical_consideration}

Generating synthetic CVs from donated real data is not exempt from risks. The main risk is the association between sensitive data (e.g., sexual orientation, ethnicity, religion) and donated CVs, which may lead to identifying a person as belonging to a minority group. In this sense, we consider internal and external risks. Regarding internal risks, researchers who have direct access to the data can associate CVs with sensitive data. The risk is mitigated by limiting access to the data on a need-to-know basis through the legal protections of a Non Disclosure Agreement. Also, data is kept encrypted and discarded after achieving the final version of the synthetic data.

Concerning the external risks, researchers accessing the synthetically generated CVs could use them to establish an association between a donated CV and sensible data. However, this risk could generally be considered minimal. First, this data intruder would need to find the original set of CVs to be able to reason about them based on the synthetic data, and neither the identity of the donors nor the corpus of CVs will be available online. Second, even if the original set of CVs were leaked, a data intruder will most likely not know which pieces of information of the synthetic CVs come from which original CVs. Hence, reasoning about the associations between identifying and sensitive information will be highly uncertain.

The procedure for collecting real CVs, the approach introduced to safeguard donors' identities, and the methods to process and generate the synthetic data have passed through a strict ethical review conducted by the Institutional Committee for Ethical Review of Projects (CIREP, by its Spanish acronym) at Pompeu Fabra University. Given the sensitive data collected during the campaign, CIREP was particularly concerned about protecting the identity of donors. After two rounds of reviews and revisions, the protocol was approved.

The dataset will be available to researchers at established European institutions, subject to a license agreement. This agreement prohibits re-identification attempts and outlines permitted uses of the corpus. Interested researchers should follow the instructions at \url{https://findhr.eu/synthetic-cvs}. According to the approved protocol, all donated materials will be destroyed following the publication of this research.


\subsection{Technical Implementation}\label{sec-tech-implementation}

The pipeline is implemented with Python and open-source data processing libraries such as Pandas, Numpy, Streamlit, Sbert, and Spacy. Bash scripts automate the execution of a series of steps. Tables and databases were developed using comma-separated values (CSV) and JSON files.

The generator was implemented through 
a command-line interface (CLI) facilitates the automatic generation of CVs in batch. The CLI application operates in two steps. The first step computes a list of plausible generation parameters by trying all different combinations of professionally related attributes (job sector and years of professional experience) and demographic attributes, such as gender, age, ethnicity, or disability condition. 

After checking whether the number of real CVs that satisfy a given combination is larger than the threshold established to safeguard donors’ identity (20 according to the generation procedure, see Section \ref{sec_synthetic_cv_generator}), the combination is included in the list of plausible generation parameters. In the second step, the CLI application takes the combination of parameters from the list of plausible generation parameters and creates synthetic CVs. A generation execution is conducted per combination. That is to say, if, for example, the list contains 50 plausible combinations of parameters, the application executes 50 attempts to generate CVs. Apart from the generated CVs---saved as JSON files---a report file is produced describing the total number of CVs generated as well as the number of CVs produced per parameters. 



\subsection{Dataset Validation}

We believe the donation dataset is valuable as a reference for studying bias in algorithmic hiring because it reflects the demographic characteristics of the European workforce. In this context, the synthetic dataset is considered valid if it demonstrates similar utility to the reference dataset. 

Our utility validation is based on the indistinguishability between the synthetic dataset and the reference dataset. In this sense, two methods are proposed. On the one hand, we conduct univariate distribution comparisons, as proposed in \cite{el2020practical}. This way, we can analyze whether variable distributions are similar in the reference and synthetic data. In particular, the distribution of the variables: job sector, years of professional experience, age, gender, ethnicity, religion/belief, and disability are computed for both datasets. Results are depicted on charts, whose shapes are visually inspected. Additionally,  Jensen–Shannon divergence \cite{nielsen2019jensen} is applied to analytically explore differences in distributions. Jensen–Shannon divergence is a technique from information theory that measures the divergence between two distributions.

On the other hand, a subjective evaluation \cite{kaloskampis2019} is conducted. In this sense, crowdsourcing workers are invited to look at 10 randomly selected CVs, 5 from the reference and 5 from the synthetic sets, and select whether they perceive the CV as real or artificially created. 
A representative random sample of 1,000 CVs, 500 from the reference and 500 from synthetic datasets, is used in this part of the evaluation, and three workers assess each CV. We make sure to include in the sample CVs corresponding to the different professional sectors 
available in the datasets. 

\begin{figure}[tb]
  \centering
  \includegraphics[width=0.8\textwidth]{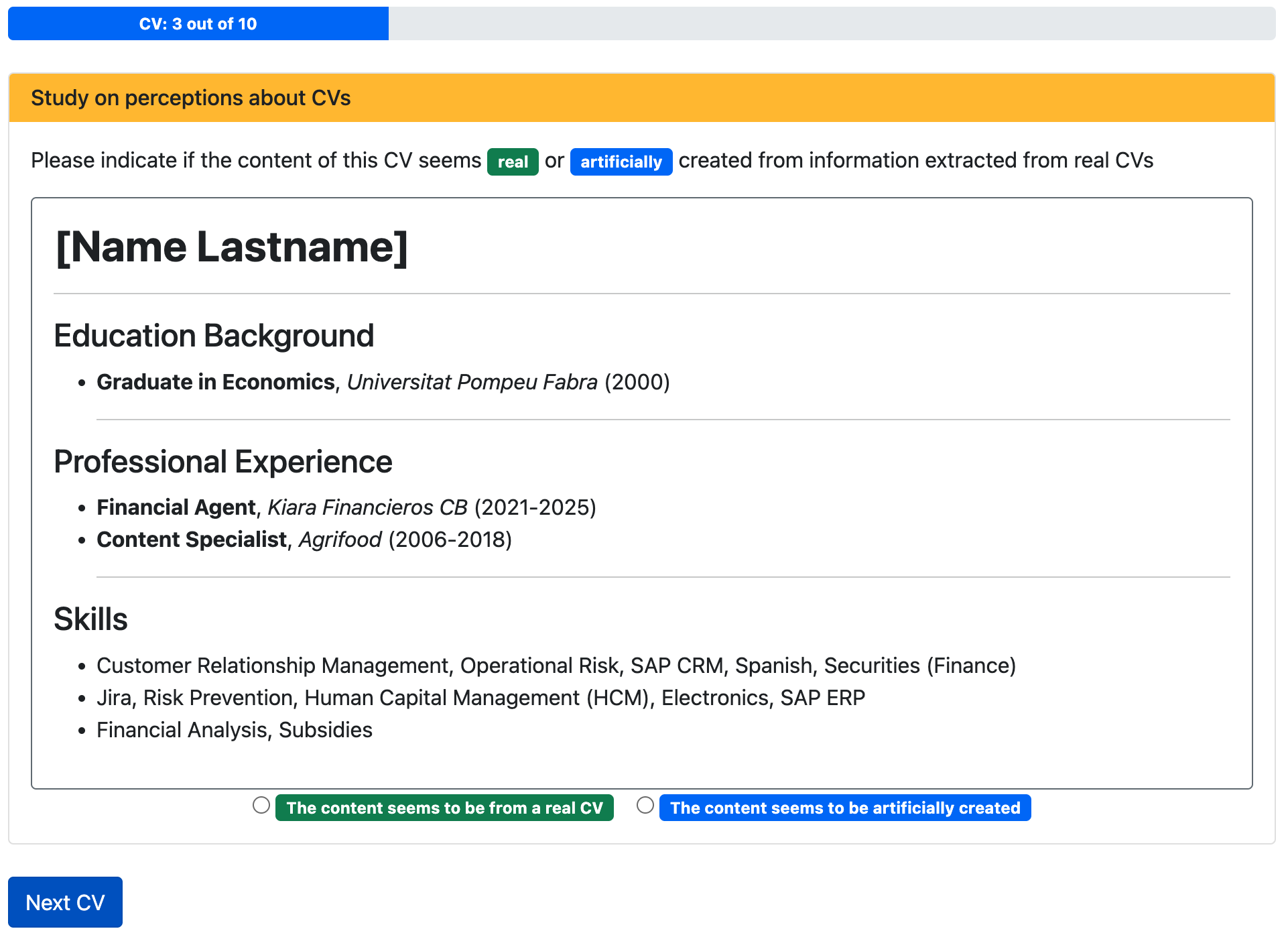}
  \caption{Screenshot of the website used in the subjective evaluation of the synthetic CVs.}
  \label{fig_cv_validation}
\end{figure}

After reading the information sheet and approving the consent form, the participants are instructed on the task with the following prompt: \textit{You will be presented with 10 CVs and asked to determine whether each one appears to be real (i.e., belonging to an actual person) or artificially generated (i.e., created from information extracted from real CVs)} and the study can be started. The 10 CVs (5 real, 5 synthetic) are presented one by one in random order and using Markdown format \cite{ovadia2014markdown}, as shown in Figure \ref{fig_cv_validation}. For each CV, the following prompt is provided: \textit{Please indicate if the content of this CV seems real or artificially created from information extracted from real CVs} and the participants are presented with the next options: \textit{The content seems to be from a real CV} or \textit{The content seems to be artificially created.} 

The subjective evaluation was piloted with 28 crowdsourcing workers using a sample of 100 CVs in total (50 real, 50 synthetic). The goal was to ensure that the instructions were clear, the options unambiguous, and the interface seamless. These pretesting sessions allowed us to correct typos in the text and adjust the mechanism that ensures that each CV is evaluated by at least three workers.

\section{Results}\label{sec:results}

The execution of the implemented approach generated a dataset of synthetic CVs, which is described next. The generation was based on 948 of the 1,211 donated CVs. The rest were excluded for various reasons, including duplication, incompatible formats, incomplete content, and excessive length. 

\subsection{Dataset of Synthetic CVs}

The initially generated synthetic dataset contained 2,218 CVs covering six job sectors, namely \textit{science and engineering}, \textit{ICT}, \textit{business and administration}, \textit{sales}, \textit{clerical support}, and \textit{legal, social, and cultural}. CVs from the initially generated dataset were manually analyzed by a research assistant (step 12 in Figure  \ref{fig_pipeline}, see Section \ref{final_steps_generation}), who evaluated the coherence of the content on a scale of 1 (poor) to 5 (excellent). CVs with an overall coherence score of at least 4 out of 5 were included in the final dataset, which comprises 1,730 documents---almost 80\% of those initially generated. The sectoral distribution of CVs is heterogeneous, with \textit{business and administration} and \textit{clerical support} accounting for over half of the dataset. Table \ref{tab_cvs_sector_distribution} shows the distribution of CVs per sector.

\begin{table}[]
\caption{Distribution of synthetic CVs per job sector}
\label{tab_cvs_sector_distribution}
\small
\begin{tabular}{p{8cm}p{2cm}p{2cm}}
\toprule
Job Sector & Number of CVs & Percentage \\ 
\midrule
Clerical support & 468 & 27.05\% \\
Business and administration & 451 & 26.07\% \\
Science and engineering & 303 & 17.51\% \\
ICT & 216 & 12.49\% \\
Sales & 208 & 12.02\% \\
Legal, social, and cultural & 84 & 4.86\% \\ 
\midrule
Total & 1,730 & 100\% \\ 
\bottomrule
\end{tabular}
\end{table}

The professional experiences reproduced in the CVs in the final dataset are predominantly clustered at the extremes of the experience spectrum. Specifically, 87\% of CVs reflect either junior (with 4 years or less of experience) or senior profiles (with 15 or more years of experience). Table \ref{tab_cvs_experience_distribution} outlines the distribution of CVs by professional experience.

\begin{table}[]
\caption{Distribution of synthetic CVs per years of professional experience}
\label{tab_cvs_experience_distribution}
\small
\begin{tabular}{p{8cm}p{2cm}p{2cm}}
\toprule
Years of Professional Experience & Number of CVs & Percentage \\ 
\midrule
4 years or less & 787 & 45.50\% \\
5-9 years & 209 & 12.16\% \\
10-14 years & 20 & 1.17\% \\
15+ years & 714 & 41.27\% \\ \hline
Total & 1,730 & 100\% \\ 
\bottomrule
\end{tabular}
\end{table}

The generated dataset reflects various demographic characteristics, including age, gender, religion, nationality, and ethnicity, but each CV highlights only one demographic attribute at a time. In other words, no synthetic CV combines multiple demographic traits, such as representing a female (gender attribute) engineer in her 40s (age attribute). Combinations involving more than one demographic parameter turned out to be impossible due to insufficient data to represent these combinations in accordance with the approved protocol, i.e., at least 20 real CVs must exist that satisfy the combination (see Section \ref{sec_content_generation}).

The synthetic CVs in final dataset generated using \textit{gender} as the demographic attribute (270) are evenly split between male and female profiles. Most synthetic CVs created with \textit{age} as the demographic factor (180) represent primarily profiles aged 40 or younger. A total of 242 CVs in the final dataset were produced using \textit{LGBTQ+} as the defining characteristic, with 6\% representing LGBTQ+ profiles. Among the 297 CVs generated with the \textit{minority} parameter, 50\% reflect profiles from minority groups. Additionally, 12\% of the CVs created with the \textit{foreign} attribute (301) represent profiles of people who feel foreign in the country where they live. Finally, synthetic CVs depict only Christian and non-disabled profiles, lacking CVs that reproduce data of individuals from other religions (e.g., Buddhism, Hinduism, Islam) and with disability conditions. 

An example of a synthetic CV of a profile with less than four years of experience in the Sales sector is shown as JSON format in Listing \ref{example-synthetic-cv}. 
Apart from the JSON format, the Synthetic CVs are also available in Markdown to facilitate their inspections.
The dataset is available for researchers in established academic institutions of the European Union (i.e., institutions having a participant number registration with the EU, known as a ``PIC'') under a free-of-charge license agreement that forbids donor re-identification attempts. Please contact the paper authors for instructions.

\begin{lstlisting}[language=json, label=example-synthetic-cv, caption={JSON example of a synthetic CV in the final dataset generated for the Sales sector}, float]
{
    "education_background": [
        {
            "institution": "UNED",
            "start_date": "April 2022",
            "end_date": "Ongoing",
            "degree": "Degree In Law"
        }
    ],
    "professional_experience": [
        {
            "institution": "Alcampo",
            "start_date": "January 2022",
            "end_date": "December 2023",
            "role": "Cashier  Stocker",
            "duration": "1 year, 11 months",
            "duration_months": 23
        }
    ],
    "skills": {
        "others": ["Literacy", "Informatics", "Social Integration", "Research"]
    }
}
\end{lstlisting}

\subsection{Validation Results}

Results of the validation are reported next. First, we discuss the findings of the univariate distribution comparisons. Later, we reflect on the subjective evaluation.

\subsubsection{Comparing Univariate Distributions} \label{univ_distributions}

Upon analyzing the distributions of variables in both the synthetic and reference datasets, we found that, overall, the distributions are fairly similar, although  we do not aim for identical distributions, as this could signal potential privacy concerns. 

\begin{figure}[tb]
  \centering
  \includegraphics[width=0.8\textwidth]{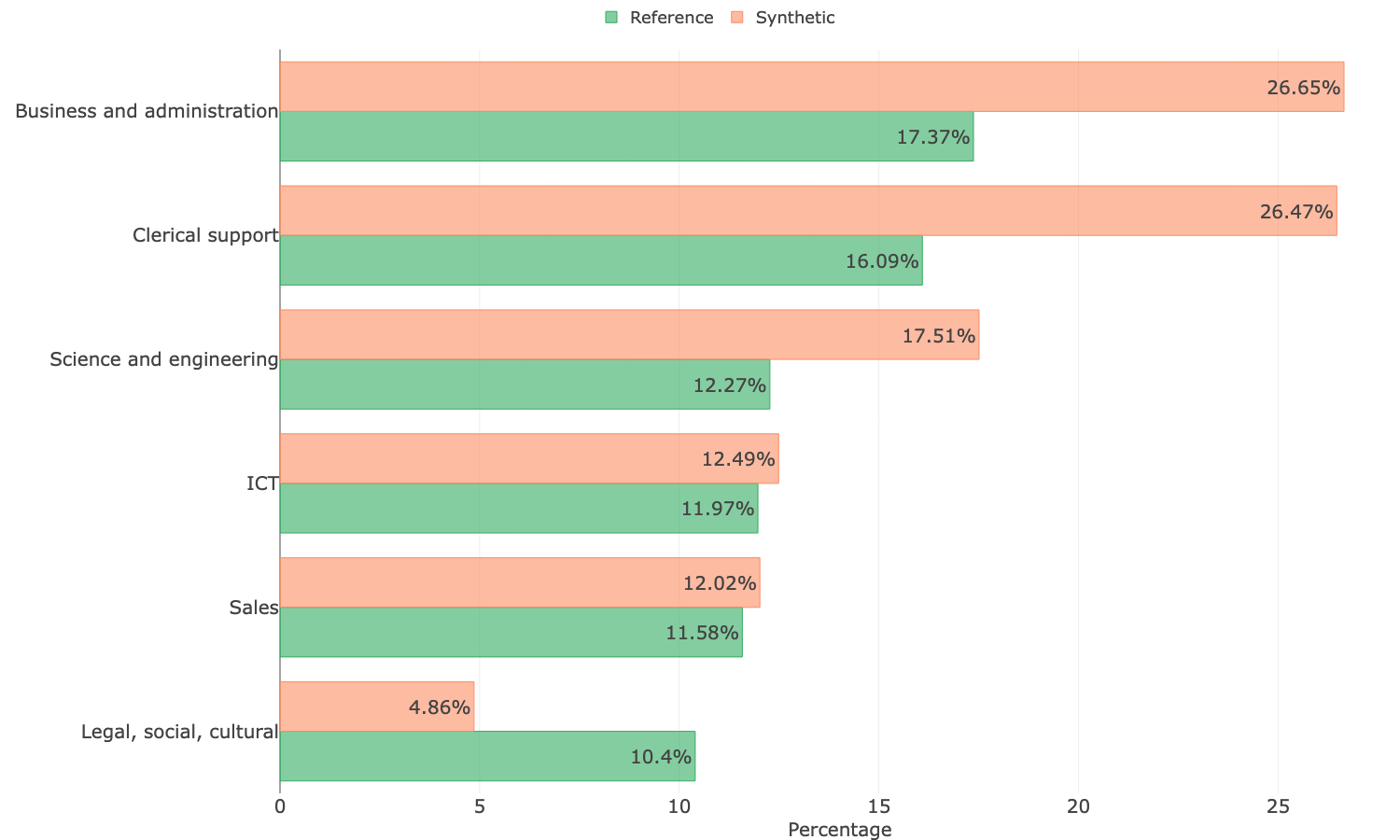}
  \caption{Comparison of job sector distributions in the reference and synthetic datasets. Values in the reference dataset do not add up to 100\% since sectors not included in the synthetic data were excluded.}
  \label{fig_job_sectors_dist}
\end{figure}

\begin{figure}[tb]
  \centering
  \includegraphics[width=0.8\textwidth]{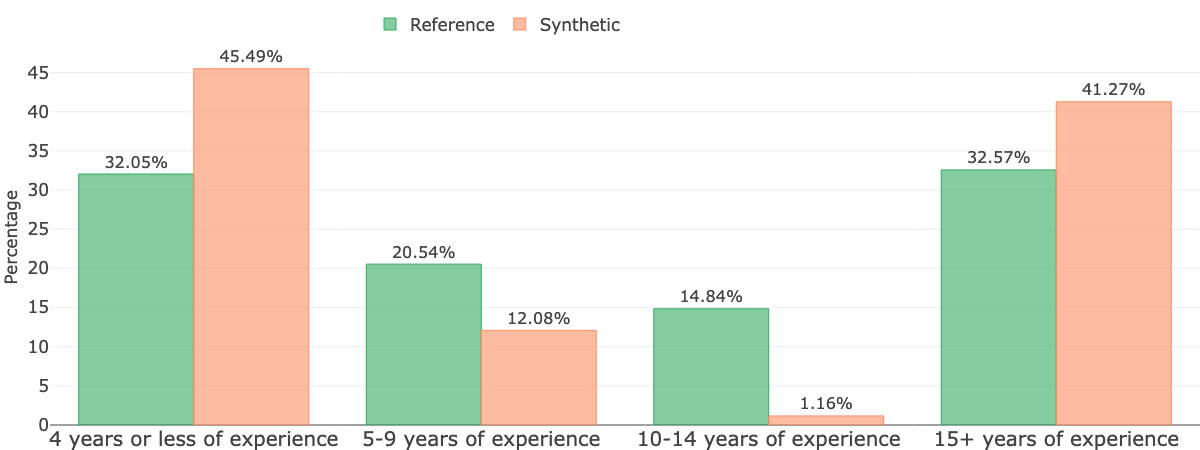}
  \caption{Comparison of years of professional experience distributions in the reference and synthetic datasets.}
  \label{fig_prof_exp_dist}
\end{figure}

As shown in Figure \ref{fig_job_sectors_dist}, the job sector distribution in the synthetic dataset is moderately close to the reference dataset, largely preserving the broad ordering of the sectors (i.e., most common to least common). 
Likewise, the distributions of years of professional experience in both datasets show to be pretty equivalent, as illustrated in Figure \ref{fig_prof_exp_dist}, suggesting a high level of data utility.

\begin{figure}[tb]
  \centering
  \includegraphics[width=0.5\textwidth]{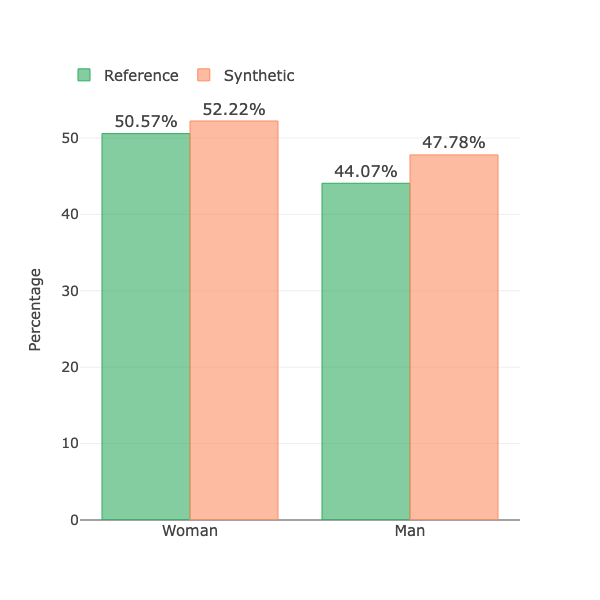}
  \caption{Comparison of gender distributions in the reference and synthetic datasets. Percentages in the reference do not up to 100\% since categories available in the reference dataset but no included in the synthetic data (e.g., non-binary) are excluded.}
  \label{fig_gender_dist}
\end{figure}

\begin{figure}
     \centering
     \begin{subfigure}[b]{0.45\textwidth}
         \centering
         \includegraphics[width=\textwidth]{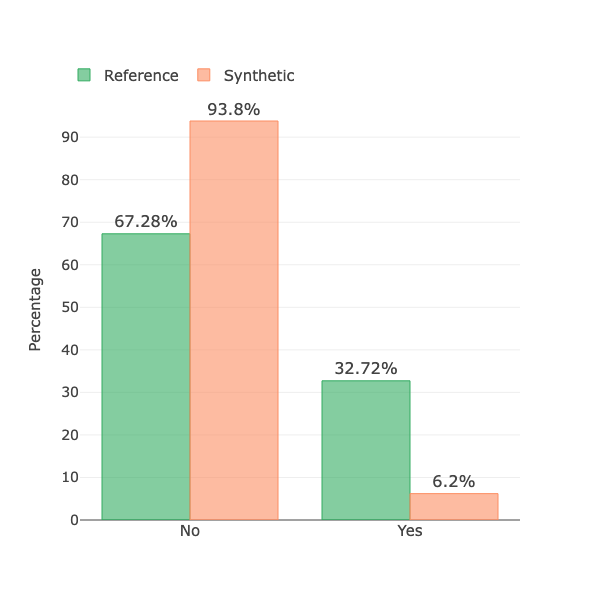}
         \caption{LGBTQ+ status distributions in the reference and synthetic datasets.}
         \label{fig_lgtbq}
     \end{subfigure}
     \hfill
     \begin{subfigure}[b]{0.45\textwidth}
         \centering
         \includegraphics[width=\textwidth]{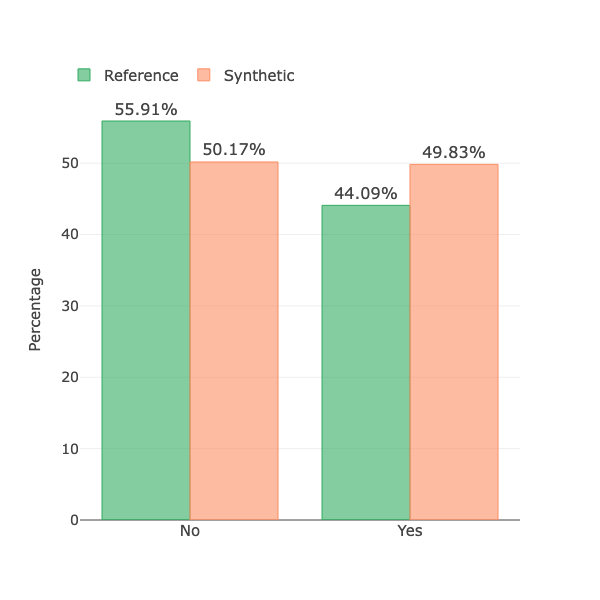}
         \caption{Ethnic minority distributions in the reference and synthetic datasets.}
         \label{fig_minority}
     \end{subfigure}
     \hfill
     \begin{subfigure}[b]{0.5\textwidth}
         \centering
         \includegraphics[width=\textwidth]{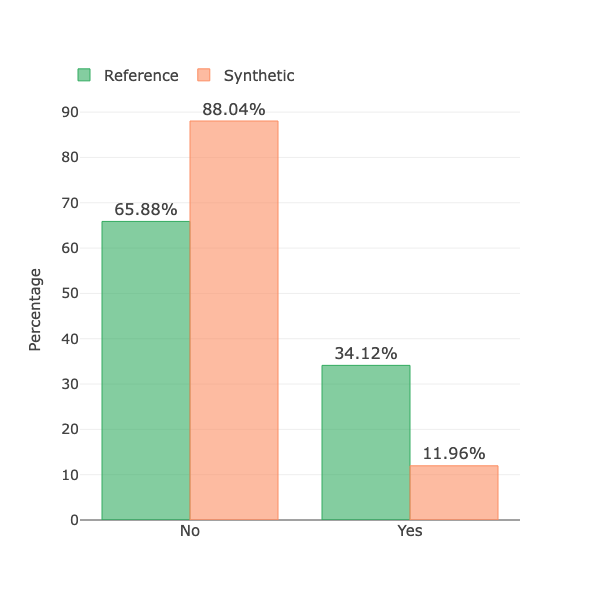}
         \caption{Perceived foreignness distributions in the reference and synthetic datasets.}
         \label{fig_foreing}
     \end{subfigure}
        \caption{Comparison of demographic attribute distributions in the reference and synthetic datasets}
        \label{fig_demo2}
\end{figure}

Regarding demographic variables, the analysis reveals that the gender distributions are identical in shape and closely balanced in terms of proportions. As shown in Figure \ref{fig_gender_dist}, the categories \textit{woman} and \textit{man} are nearly evenly distributed across both datasets. When examining age distributions, we observe that, as in previous cases, the overall ordering is well preserved. 
The distribution of categories for LGBTQ+, ethnic minority status, and perceived foreignness in the reference and synthetic datasets exhibits a nearly identical pattern in which the general ordering is conserved, as shown in Figure \ref{fig_demo2}. \textcolor{black}{The ratio between the majority and minority classes is not always aligned because limitations in the reference dataset force the generator to under-/over represent them.} 

\begin{figure}[tb]
  \centering
  \includegraphics[width=0.7\textwidth]{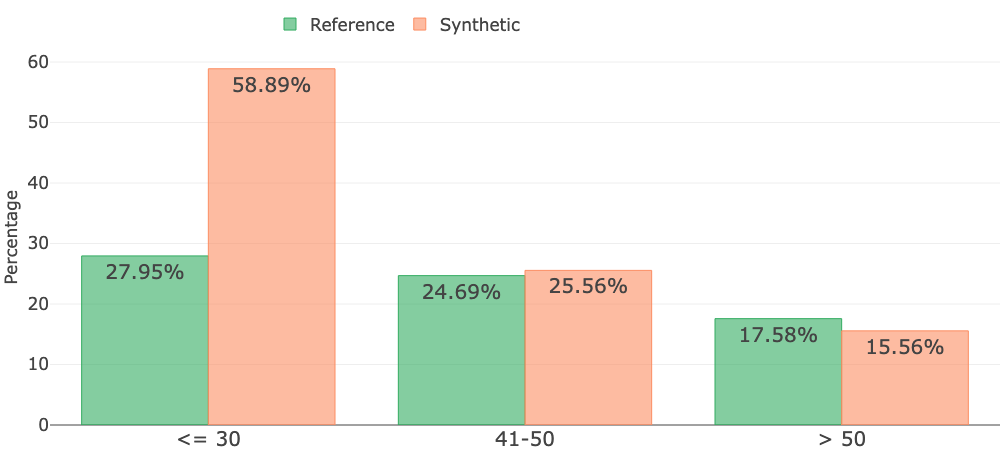}
  \caption{Comparison of age distributions in the reference and synthetic datasets. Only categories available in both datasets are included.}
  \label{fig_age_dist}
\end{figure}

To complement the visual inspection of distributions across variables, we analytically explored how distributions in the reference dataset differ from the synthetic ones. The difference between these distributions is calculated through the Jensen-Shannon divergence technique, which account for missing categories as in some cases (e.g., job sector, gender, age) and provides a numerical score between 0 and 1 representing how similar the distributions are from one another. The closer the score closer to 0, the perfect align between the distributions.

Table \ref{tab_jsscores} reports the Jensen-Shannon scores obtained for the distributions of the reference and synthetic datasets across variables. As shown, the scores are generally close to 0, with distribution similarities ranging from almost perfect---such as for \textit{gender} and \textit{minority}---to slightly different, as in \textit{LGBTQ+} and \textit{years of professional experience}.

\begin{table}[]
\begin{threeparttable}
\caption{Jensen-Shannon (JS) scores for distribution comparison across variables.}
\label{tab_jsscores}
\footnotesize
\begin{tabular}{p{1cm}p{1.5cm}p{3cm}p{1cm}p{1cm}p{1.5cm}p{1cm}p{1.2cm}}
\toprule
 & Job sector & Professional Experience & Gender & Age & LGBTQ+ & Minority & Foreigness \\ 
 \midrule
JS score & 0.12 & 0.22 & 0.01 & 0.14 & 0.25 & 0.04 & 0.20 \\ 
\bottomrule
\end{tabular}
    \begin{tablenotes}
      \footnotesize
      \item \textbf{Note.} The closer the score to 0, the similar the distributions. 
    \end{tablenotes}
\end{threeparttable}
\end{table}

In summary, our analysis reveals similar patterns between the reference and generated distributions across variables. These findings suggest that the synthetic CVs are likely to be as useful as the real documents for studying bias in algorithmic hiring, while not revealing personal information. As previously discussed, we understand that the collected CVs possess the necessary characteristics for this purpose.

\subsubsection{Subjective Evaluation} \label{subj_eval}

In total, 300 crowdsourcing workers participated in the subjective evaluation of the synthetic CVs, each conducting a single run of the experiment, i.e., no worker repeated the study. All participants declared fluency in English, half identified themselves as women, and the large majority (70\%) aged 20-40 years.

Of the 300 submissions, 14 were discarded due to low-quality answers or exceptionally high-speed execution (i.e., less than 180 seconds when the median completion time is almost 7 minutes).
On average, real and synthetic CVs were evaluated 2.86 times. The participants’ general accuracy in correctly identifying whether a CV was real or synthetic was 53\%. The probability of correctly classifying a real CV was 0.52, while the probability of a synthetic CV being mistaken for a real one was 0.55. 
Moreover, almost 90\% of the synthetic CVs (442 out of 500) were classified as real for at least one of the three evaluators, and more than half of them (260 out of 500) for two evaluators.

\section{Discussion}\label{sec:discussion}


The quantitative results 
reveal that distributions across variables in both datasets (reference and synthetic) show similar patterns, suggesting high utility of the synthetic CVs to study bias and benchmark mitigation techniques in algorithmic hiring.

Additionally, the results of the qualitative analysis of the synthetic CVs highlight their quality, as they appeared convincingly real more than half the time, demonstrating that they are not easily distinguishable as artificially created. Moreover, 90\% of them were perceived as belonging to a real person by at least one of the evaluators, reinforcing their realistic appearance.

On the other hand, we can see that the synthetic dataset is in some way skewed towards sectors that are usually dominated by professionals with, at minimum, college studies (e.g., \textit{business and administration}, \textit{ICT},  \textit{science and engineering}). Sectors---available in donations---whose workers learn occupation skills primarily from practical experiences rather than formal education (e.g., \textit{construction and manufacturing}, \textit{transport}, \textit{personal service}), are not represented in the synthetic dataset due to their low representativeness in the reference dataset, which avoids the generator to apply specific combination of parameters that do not satisfy the requirements of the protocol set in place to safeguard donors' identities. 

The same applies to some demographic characteristics missing in the synthetic dataset, which are a result of insufficient data in the reference dataset, such that the generator could operate while fulfilling the protection measures proposed to safeguard donors' identities. Furthermore, generating CVs with combinations involving more than one demographic parameter proved not plausible for the same reason.

\textcolor{black}{Beyond these characteristics, our dataset shares certain similarities with, but also presents significant differences from, the collections of CVs described in Section~\ref{sec:related_works} (see Table \ref{tab_previous-datasets} for a summary). In terms of similarities, it includes synthetically generated CVs, as in \cite{skondras2023generating, pena2020bias, bruera2022generating}, and, like  \cite{skondras2023generating,kaggle_dataset,bruera2022generating,de2019bias}, it spans multiple professional sectors. Its content also resembles that of most reviewed datasets, with the exception of \cite{skondras2023generating,jiechieu2021skills,de2019bias}.}

\textcolor{black}{As for the differences, our dataset is, among the reviewed literature, the only synthetic collection created through a process explicitly designed to safeguard personal information and grounded in real CVs gathered specifically for this purpose. By contrast, previous efforts (e.g., \cite{skondras2023generating, kaggle_dataset, de2019bias}) have largely relied on web-scraped content. Moreover, our dataset introduces sensitive demographic attributes not previously available to researchers (e.g., disability condition, LGBTQ+ status). Finally, unlike earlier synthetic or semi-synthetic datasets (e.g., \cite{RecSysChallenge2016_TrainingDataset, RecSysChallenge2017_Dataset, bruera2022generating}), every generated CV underwent manual review. Apart from \cite{kaggle_dataset, drushchak-romanyshyn-2024-introducing}, which are publicly available, our dataset is accessible for researchers in established academic institutions within the European Union, i.e., they are eligible to receive research funding from the European Commission.} 


\subsection{Limitations and Future Work}

The generation approach and the synthetic dataset have limitations.
Next, we present details of these limitations and propose alternatives to address them in future work.

\subsubsection{Generation Approach}

Even when donations in formats different from PDF (i.e., DOC/DOCX, ODT) were received, the generator did not process them, and they were ignored. Additionally, due to limitations in the document parser, CV documents with more than 10 pages are not considered. These limitations prevented us to use the entire pool of donated CVs. Also, donations that do not include attached CV documents but the information about the donor’s educational background, professional experiences, and skills are shared in the donation form, are omitted.

CVs not written in English are automatically translated into English with the risk of losing the nuances and particularities of the original language. 
The generator expects to operate on a fixed number and type of data fields, and it is compatible with a pre-defined structure of parsing output. Moreover, although it is built on free and open-source libraries and software, it depends on a third-party paid parsing service. Albeit replacing text processing dependencies, like the parser, does not represent a significant change, and can be done relatively easily. 

As mentioned in Section \ref{sec:method}, hand-made rules have been included in the generator to guarantee certain quality and consistency in the results. Through the various iterations of the CV generator, rules have been modified and added to improve the generation quality. However, we acknowledge that producing an exhaustive set of rules that covers all possible situations is unrealistic; therefore, unexpected inconsistencies in the content of the synthetic CVs might be present. To overcome this challenge and ensure the quality of the final dataset, a manual validation step has been included in the pipeline.

Additionally, we aim to extend the generator to include a module that looks for plausible substitutes for academic institutions and workplaces. So, instead of using the academic institutions or workplaces originally found in the donated CVs, a plausible substitute will be identified and employed in the creation of the synthetic CV. For example, instead of using the University of Pisa in an education item of a synthetic CV (assuming that this university was included in some of the donated CVs), we could use the University of Trento, as those institutions come from the same country and are similar in terms of size and education offers. This feature will further enhance the measures adopted to protect donors' identities.

\subsubsection{Synthetic dataset}

As we saw, the generator reproduces the distribution of the collected data. Hence, if it operates on a biased source dataset those biases will be replicated in the synthetic dataset as no artificial corrective actions (e.g., up-sampling) have been performed to balance the seed dataset. Similarly, it was impossible to generate CVs for sectors without representation in the donated data. 
To address this issue, we propose to create CVs for some of the underrepresented sectors. In particular, we have already performed online research and evaluated the usual characteristics of CVs from the \textit{education}, \textit{tourism} and \textit{arts \& crafts} sectors, and plan to manually generate 100 artificial CVs for each of these sectors.

\section{Conclusions}\label{sec:conclusion}

According to the AI Act, AI systems intended to be used for the recruitment or selection of natural persons, in particular to place targeted job advertisements, to analyze and filter job applications, and to evaluate candidates, are considered high-risk, as these systems could significantly affect individuals' fundamental rights. Therefore, the development of anti-discriminatory methods and algorithms contextualized within the actual technical, legal, and ethical context of algorithmic hiring is of utmost importance. 

To advance research in this field, given the current lack of datasets reflecting the diverse characteristics found in real CVs of the actual workforce, in this work we present a dataset of 1,730 synthetic CVs representing six work sectors, along with the approach designed to generate them.

The developed dataset follows the AI Act recommendations on the use of synthetic data as a primary option to avoid processing sensitive personal data, when aiming to detect and mitigate bias \cite{council2024regulation}. Its envisioned use is as a benchmark during the pre-deployment phase of candidate ranking systems developed with relevance as a goal, to detect whether those systems result in discriminatory outcomes, based on attributes such as age, gender, or nationality, disadvantaging systematically some groups of applicants. 

The dataset was developed using real CVs, collected through a data donation campaign, and found to follow the features' distribution in the donated data. Its delivery and use will permit an increased research reproducibility, while the proposed generation technique supports an increased explainability of the outcomes. Therefore, this work is expected to have an important impact on both the industrial and academic sectors.

Finally, this work is expected to be a significant contribution to current standardization efforts, especially in the direction of \textit{Data Governance and Quality}. We envision that the presented synthetic dataset can become a standard for fairness benchmarking of algorithmic hiring systems, especially when implementing AI systems for candidate ranking based on CVs, as our dataset enables the comparison of multiple ranking algorithms, to ensure they do not introduce or reproduce discriminatory behaviors based on candidates' protected characteristics. In addition, in cases where the specific dataset cannot be directly used, for instance, when a different working force population is evaluated, the proposed methodology allows the generation of an adequate benchmarking dataset.

\begin{acks}
This work was partially supported by the project FINDHR (Horizon Europe research and innovation program, grant no.: 101070212). The authors would like to thank Marta Gracia, especially for her collaboration in processing the donated CVs and Tamara Vorobyeva for the careful evaluation of the generated CVs.
\end{acks}

\bibliographystyle{ACM-Reference-Format}
\bibliography{0_main}

\end{document}